\documentstyle[amssym]{mn}
\input{epsf.sty}

 at 10truept

\def\and  {\it {et al.} \rm}

\def\etal{{\rm et~al. }}
\def\spose#1{\hbox to 0pt{#1\hss}}
\def\simlt{\mathrel{\spose{\lower 3pt\hbox{$\mathchar"218$}}
     \raise 2.0pt\hbox{$\mathchar"13C$}}}
\def\simgt{\mathrel{\spose{\lower 3pt\hbox{$\mathchar"218$}}
     \raise 2.0pt\hbox{$\mathchar"13E$}}}
\def\beq{\begin{equation}}
\def\eeq{\end{equation}}
\def\bce{\begin{center}}
\def\ece{\end{center}}
\def\bea{\begin{eqnarray}}
\def\eea{\end{eqnarray}}
\def\ben{\begin{enumerate}}
\def\een{\end{enumerate}}

\def\brr{\begin{array}}
\def\err{\end{array}}

\def\etal{{\rm et~al. }}

\def\nh1{n_{\rm HI}}

\def\p1dk{P_{\rm 1D}(k)}
\def\simlt{\mathrel{\spose{\lower 3pt\hbox{$\mathchar"218$}}
     \raise 2.0pt\hbox{$\mathchar"13C$}}}
\def\simgt{\mathrel{\spose{\lower 3pt\hbox{$\mathchar"218$}}
     \raise 2.0pt\hbox{$\mathchar"13E$}}}

\def \be {\begin{equation}}
\def \en {\end{equation}}
\def \bea {\begin{eqnarray}}
\def \ena {\end{eqnarray}}

\def \bi {\begin{itemize}}
\def \ei {\end{itemize}}

\def \eg {{\it e.g. }}
\def \ie {{\it i.e. }}
\def \etal {{\it et al. }}
\def \prd {{\it Phys. Rev. D}}
\def \prl {{\it Phys. Rev. Let.}}
\def \apj {{\it ApJ}}
\def \apjl {{\it ApJ Let.}}

\def \aa {{\it A\&A}}
\def \mnras {{\it MNRAS}}
\def \nat {{\it Nature}}

\begin{document}

\title[Probing CMB Non-Gaussianity Using Local Curvature]{Probing CMB
Non-Gaussianity Using Local Curvature}

\author[ O. Dor{\'e}, S. Colombi \& F.R. Bouchet]
        {Olivier Dor{\'e}, St{\'e}phane Colombi and Fran\c cois R. Bouchet\\
        \emph{Institut d'Astrophysique de Paris}, 98bis boulevard
        Arago, 75014 Paris, FRANCE\\
        dore@iap.fr, colombi@iap.fr, bouchet@iap.fr}

\maketitle

\begin{abstract}
It is possible to classify pixels of a smoothed cosmic microwave
background (CMB) fluctuation map according to their
local curvature in ``hill'', ``lake'' and ``saddle'' regions. 
In the Gaussian case, fractional areas occupied by 
pixels of each kind can be computed analytically  
for families of excursion sets as functions of threshold 
and moments of the fluctuation power spectrum. 
We show how the shape of these functions can be used to constrain
accurately the level of non-Gaussianity in the data by applying 
these new statistics to an hypothetical mixed model suggested 
by Bouchet \etal (2001). According to our simple test, with only one $12.5 \times 
12.5\ \mathrm{deg}^2$ map, Planck should be able to  detect with a
high significance a non-Gaussian level as weak as $10\%$ in
temperature standard deviation (rms) (5\% in $C_{\ell}$), whereas a marginal detection would be
possible for MAP with a non-Gaussian level around $30\%$ in
temperature (15\% in $C_{\ell}$). 
\end{abstract}

\section{Introduction}

From the now well measured temperature fluctuations in the
cosmological microwave background (CMB) we can gain some
unvaluable constraints on the physics of the early universe. An
important issue lies in determining whether these fluctuations are of
Gaussian nature or not. Indeed, most of inflation scenarios
predict a very low level of non-Gaussianity while models
involving topological defects can give rise to significantly
non Gaussian fluctuations. Even if a high degree of non-Gaussianity
seems disfavored by recents measurements
\cite{Net01,Lee01,Hal01,Sa01,PoAd02}, the presence of topological
defects at a significant level is
not yet ruled out by observations, as advocated recently by
e.g. Bouchet \etal (2000), who considered a mixed model where Cosmic
Microwave Background (CMB) temperature fluctuations are seeded in part by cosmic strings. 

There are numerous ways of probing non Gaussian features of a random
field such as a CMB temperature fluctuations map. 
Since statistical properties of random Gaussian 
fields are entirely determined by their two-point correlation function,  
a natural approach consists in measuring higher order
correlation functions or related statistics such as cumulants
of the distribution. The measurements can be done in real and Fourier
space \cite{Hi95,FeMa98,Ma00,BaZa00,GaMa00a,GaMa00b,VeHe01,Ko01,Sa01,KuBa01} or in the
space of wavelet coefficients \cite{AgFo99,AgFo01}. 

Alternatively, non Gaussian features of CMB maps can be 
probed by the topological analysis their excursions sets, defined for
a random field $T({\mathbf \theta})$ as $A_u(T) = \{{\mathbf \theta}\ |\ T({\mathbf \theta}) \ge u
\}$.  For example, for a sufficiently smooth and non degenerate
random field it is possible to measure the Euler
characteristic (or genus) of the excursion sets via a counting of critical 
points\footnote{The points where the gradient of the field cancels.},
classified according to their curvature, \ie maxima, minima or
saddle points in our 2D case (e.g. Adler 1981, p.~87). 
Critical points counting is a well known
practical issue in the context of CMB analysis and 
specific predictions can be derived for smooth random Gaussian fields (e.g., 
Adler 1981), which allow one to constrain the degree of non
Gaussianity of CMB maps by measuring e.g. the Euler characteristic 
\cite{Co89,GoPa90,Luo94,Sm94,BaMa01} or peak statistics \cite{BoEf87,Co89,NoJo96,He99,He01}. 

\begin{figure*}
\centering
\centerline{\epsfxsize=12cm \epsfbox{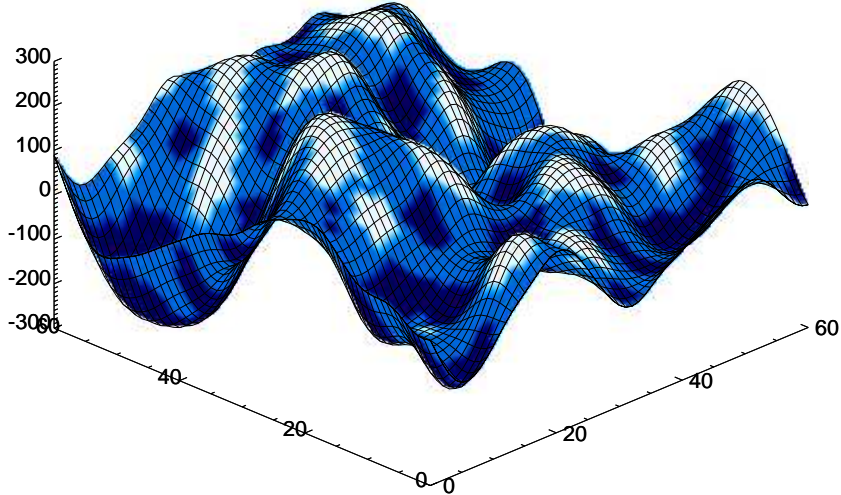}}
\caption{Example of location of ``lake'',  ``hill'' and ``saddle'' points in
a Gaussian random field smoothed with a Gaussian window of size
5 pixels. The points are colored according to their
type: ``hill'' points are in white, ``lake'' points in dark grey and
``saddle'' points in light grey. Lakes are connected to light grey
hills by dark grey saddles. ``Lake'' and ``hill''points
are of comparable abundance while ``saddle'' points are the most
common. The evolution of the relative abundances
with thresholding can be well visualized from this plot: for example, when the
threshold gets higher, the relative abundance of lake points decreases, 
while on the contrary the fraction of hill points increases, as
expected. In this paper we shall rely on  
details of these variations to discriminate between a Gaussian and
a non Gaussian random field.
\label{plot_3d_type_2}}
\end{figure*}

Following an approach advocated recently in the large scale structure 
context (3D) by Colombi, Pogosyan \& Souradeep (2000), we propose in this paper 
to extend the counting to ordinary points, \ie to classify {\em all} the points according
to their local curvature as belonging to ``hills'',  ``lakes'' or
``saddles'' (see Fig.~\ref{plot_3d_type_2} for an example).  
By measuring the relative abundances, ${\cal P}_{\rm hill}$, ${\cal P}_{\rm lake}$ and ${\cal
P}_{\rm saddle}$, of these three types of points for various excursions
sets and smoothing scales, \ie by exploring the correlation between
height and curvature\footnote{In this sense, our work is similar in
spirit as in Takada (2001).}, we are able to extract a mathematically
well defined Gaussian signature, depending formally on only one specific
ratio of spectral parameters. As a result, a comparison of the measured abundances with
those predicted in the Gaussian case allows us to detect a certain level of 
non-Gaussianity.

This paper is organized as follows. In section {\S}~\ref{theory}, after
recalling some useful results  
about 2D Gaussian random fields, we derive analytic predictions 
for the abundances, by extending the work of 
Bond \& Efstathiou (1987). In section {\S}~\ref{measure} we discuss
technical issues involved in the practical measurements of ${\cal
P}_{\rm hill}$, ${\cal P}_{\rm lake}$ and ${\cal P}_{\rm saddle}$. In
section {\S}~\ref{ngtest} we build   a $\chi^2$ statistic that allows us
to combine a set of measurements in order to quantify the likelihood
of an image to be Gaussian, from the standpoint of this measure. 
We then apply the method to the case of
noisy mixed models  involving cosmic strings plus cold dark
matter. Finally, results are discussed in section {\S}~\ref{discuss}.


\section{Theoretical Considerations}
\label{theory}

In this section we recall some specific properties of 2D Gaussian
random fields, following Bond \& Efstathiou (1987) (BE87) (a 2D transcription of
the 3D formalism of Bardeen \etal 1986). We extend the calculations
of this work to obtain theoretical expressions for the abundances
of interest, i.e. the fractions of space occupied 
respectively by hill, lake and saddle regions. 

\subsection{Facts about a Gaussian random field}

Let us first consider a temperature fluctuation field,
$\delta_{\rm T}({\mathbf r}) = T({\mathbf r})/\bar{T}-1$ whose $2D$ Fourier
transform is $\delta_{\rm T}({\mathbf k}) \equiv \int d^2{\mathbf
k}\delta_{\rm T}({\mathbf r})\exp(i{\mathbf k.r})$ and whose
power spectrum $P(k)$ is
\be
\langle \delta_{\rm T}({\mathbf k}) \delta_{\rm T}({\mathbf k'}) \rangle =
(2\pi)^2P({\mathbf k})\delta_D({\mathbf k}+{\mathbf k'}),
\en 
where $\delta_D({\mathbf k})$ is the Dirac distribution. 
We can define the moments of the power spectrum 
\be
\sigma_j^2 \equiv {1 \over (2\pi)^2}\int d^2{\mathbf k}\ P(k)k^{2j}
\label{sig2j}
\en 
and the ratio 
\be
\gamma \equiv {\sigma_1^2 \over \sigma_0 \sigma_2}.
\label{gammadef}
\en
Note that $\sigma_0$ is the rms of the fluctuation field
$\delta_{\rm T}$. Naturally, in the flat sky approximation, \ie large
multipole $l$ and small separation angles, the Fourier (flat) power spectrum  is related
to spherical harmonic power spectrum as follows 
\be
P(k) \simeq C_{\ell} \quad \mathrm{and} \quad k \simeq \ell.
\en

Let us consider from now on only the normalized fluctuation field
\be
\delta(\mathbf{x}) \equiv {\delta_{\rm T}(\mathbf{x}) \over \sigma_0} .
\en 
At a given point ${\mathbf r}$, 
we note the gradient of the field $\nabla \delta
\equiv {\mathbf \eta}$ and the Hessian matrix, $\zeta_{ij} \equiv
\partial \delta/\partial x_i\partial x_j$. This
symmetric real matrix can be diagonalised by applying a rotation of an
angle $\theta$ and we note $\lambda_1$ and $\lambda_2$ the
\emph{opposite} of its eigenvalues, with $\lambda_1 \ge
\lambda_2$. The normalized trace of the curvature matrix reads,  
\be
x \equiv {\lambda_1+\lambda_2 \over \sigma_2},
\en
and the ellipticity $e$ is defined as 
\be
e={\lambda_1-\lambda_2 \over 2\sigma_2 x}.
\en 
Note that with this notation, \emph{$x$ and $e$ should have the same
sign}. If $\lambda_2\ge 0$ then $x\ge 0$ and $0\le e \le 1/2$, if
$\lambda_1\le 0$ then $x\le 0$ and $-1/2\le e \le 0$. If $\lambda_1$
and $\lambda_2$ have opposite signs, then neither $x$ nor $e$ are restricted.

For a Gaussian random field, the joint  probability distribution function of $\delta$,
${\mathbf \eta}$, $x$, $e$ and $\theta$ is given by [Eq.~(A1.6) in BE87]
\bea
\lefteqn{{\mathcal P}(\delta,{\mathbf \eta},x,e,\theta)d\delta\ dx\ de\ d\theta\ d^2 {\mathbf \eta} = } 
\nonumber     \\
& & e^{-\delta^2/2}{\mu d\delta \over \sqrt{2\pi }} e^{-\mu ^2(x-\gamma\delta)^2/2}{dx \over
\sqrt{2\pi}}\ \times \nonumber\\ 
& & e^{-{\mathbf \eta}^2/\sigma_1^2} {d^2 {\mathbf \eta} \over
\pi\sigma_1^2} e^{-4x^2e^2}8x^2ede{d\theta \over \pi},
\label{p_gauss}
\ena
where we set 
\be
\mu \equiv (1-\gamma^2)^{-1/2}.
\en 
Knowing this probability distribution function, it is easy to show that
$\langle \delta^2 \rangle=\langle x^2 \rangle=\langle {\mathbf \eta}^2
\rangle=1$, that ${\mathbf \eta}$ is
correlated with neither $\delta$ nor $x$ nor $e$, but that $\langle \delta x \rangle = \gamma$, 
\ie the height of a point is correlated with the curvature at this
point. This lattest result is of particular interest to us since the
so-induced non-trivial dependence on thresholding will generate a specific gaussian
signature that we will exploit in the following. 

\subsection{Studying the local curvature}

BE87 calculated the density probability of
extrema for a Gaussian random field as a function of threshold. So
using the properties of \emph{first derivatives}, they open the way to
further works \cite{He99,He01} that illustrate the use of the
number and spatial distribution of extrema to characterize Gaussian random
fields.

We aim at extending this work to \emph{second derivatives} by
characterizing the local curvature \emph{at any point} and then
by comparing the abundance of $3$ defined types.

The local curvature is defined by the Hessian. By considering the sign
of its eigenvalues ($-\lambda_1$ and $-\lambda_2$) we are led to
distinguish three families of points~:
\bi
\item the ``hill'' points for which both eigenvalues are
negative, \ie $x \ge 0$ and $0 \le e \le 1/2$;
\item the ``lake'' points for which both eigenvalues are
positive, \ie $x \le0$ and $-1/2 \le e \le 0$;
\item the ``saddle'' points for which the eigenvalues are of different signs.
\ei
The first, second and third family incorporate respectively maxima, minima and saddle points.

Extending the calculation of BE87 we now compute the probability
that a point ($\delta,{\mathbf \eta},x,e,\theta)$ above a given
threshold  $\delta_{\rm th}$, \ie such that $\delta \ge \delta_{\rm th}$,
belongs to any of these three classes. More precisely we calculate
the quantities ${\mathcal P}_{\rm hill}(\delta_{\rm th}) \equiv
{\mathcal P}(0\le x,0\le e\le 1/2 | \delta_{\rm th}\le \delta)$,
${\mathcal P}_{\rm lake}(\delta_{\rm th}) \equiv {\mathcal P}(0\ge x,0\ge e\ge
1/2 | \delta_{\rm th}\le \delta)$ and ${\mathcal P}_{\rm saddle}(\delta_{\rm th}) =
1 - {\mathcal P}_{\rm hill}(\delta_{\rm th}) -
{\mathcal P}_{\rm lake}(\delta_{\rm th})$.\\

To compute the quantity ${\mathcal P}_{\rm hill}(\delta_{\rm th})$ we first
estimate the probability ${\mathcal P}(0\le x,0\le
e\le1/2,\delta_{\rm th}\le \delta)$.
Starting from the distribution, Eq.~(\ref{p_gauss}), for the cases of
interest to us, we can perform straightforwardly both the integration
$\int_{-\infty}^{\infty}d^2 {\mathbf \eta}/ (\pi\sigma_1^2)
\displaystyle$ and $\int_{0}^{\pi}d\theta / \pi \displaystyle $. Doing
so yields a  probability function ${\mathcal P}(\delta,x,e)\displaystyle$. 
Note that the fact that $\theta \in [0,\pi]$ is due to the chosen
ordering of the eigenvalues ($\lambda_1 \ge \lambda_2$) implicit in the
distribution Eq.~(\ref{p_gauss}). The subsequent integration $
\int_0^{1/2} de\ \ \displaystyle$ yields the differential density  
\bea
\lefteqn{{\mathcal N}_{\rm hill}(\delta,x,0\le e\le 1/2)\ d\delta dx\
=}\nonumber       \\
&& \ e^{-\delta^2/2}\ {\mu d\delta \over (2\pi)^{1/2}}
e^{-\mu ^2(x-\gamma\delta)^2/2} (1-e^{-x^2}) {dx \over(2\pi)^{1/2}} .
\ena
Then we can still perform analytically the integration over $x \ge 0$ and we get
\bea
\lefteqn{{\mathcal N}_{\rm hill}(\delta,0 \leq x,0\le e\le 1/2)d\delta\ \ = }\nonumber       \\
&& e^{-\delta^2/2}\ {\mu d\delta \over 2 \sqrt{2\pi}}\ \Bigl[\ {1 \over \mu }
\ \Bigl( 1 + \mathrm{Erf}\Bigl({\gamma \mu \delta \over
\sqrt{2}}\Bigr)\ \Bigl) - \nonumber \\
&& {e^{-\mu ^2\gamma^2\delta^2/(2+\mu ^2)} \over \sqrt{2+\mu ^2}} \Bigr(1 + \mathrm{Erf}\Bigl({\gamma
\mu ^2\delta \over \sqrt{2}\sqrt{2+\mu ^2}}\Bigr)\ \Bigr)\ \Bigr]  .
\ena
The integration over some threshold  $\delta_{\rm th}$ cannot be performed
analytically. However, knowing that
\be
{\mathcal P}(\delta_{\rm th}\le \delta) = {1 \over 2} \mathrm{Erfc}\left( {\delta_{\rm th}
\over \sqrt{2}}\right)\; ,
\en
it is easy to evaluate numerically the quantity
\be
{\mathcal P}_{\rm hill}(\delta_{\rm th}) = {\int_{\delta_{\rm th}}^{\infty}\ {\mathcal N}_{\rm hill}(\delta,0 \leq x,0\le e\le
1/2)d\delta \over {\mathcal P}(\delta_{\rm th}\le \delta)}.
\label{plake_num}
\en
We can still obtain analytically the limiting value
${\mathcal P}_{\rm hill}(\delta_{\rm th}=-\infty)$, \ie the fraction of
``hill'' points in the absence of threshold. We find
\be
{\mathcal P}_{\rm hill}(\delta_{\rm th}=-\infty) = {1 \over 2}\left(1-{1\over
\sqrt{3}}\right) \simeq 0.2113\; .
\en

An analogous calculation can be performed for ``lake'' points. It
leads to the differential density
\bea
\lefteqn{{\mathcal N}_{\rm lake}(\delta,x \leq 0,-1/2\le e\le 0)d\delta\ \
= }\nonumber          \\
&& e^{-\delta^2/2}\ {\mu d\delta \over 2 \sqrt{2\pi}}\ \Bigl[\ {1 \over \mu }
\ \mathrm{Erfc}\Bigl({\gamma \mu \delta \over \sqrt{2}}\ \Bigl)\ -\
\nonumber \\
& &{e^{-\mu ^2\gamma^2\delta^2/(2+\mu ^2)} \over \sqrt{2+\mu ^2}}
\mathrm{Erfc}\Bigl({\gamma \mu ^2\delta \over
\sqrt{2}\sqrt{2+\mu ^2}}\Bigr)\ \Bigr] 
\ena
from which we can deduce numerically ${\mathcal P}_{\rm lake}(\delta_{\rm th})$
as in Eq.~(\ref{plake_num}). The asymptotic value 
${\mathcal P}_{\rm lake}(\delta_{\rm th}=-\infty)$ can also be
obtained analytically, and from parity argument one finds
\bea
{\mathcal P}_{\rm lake}(\delta_{\rm th}=-\infty) & = & {\mathcal
P}_{\rm hill}(\delta_{\rm th} =-\infty) \\
& = & {1 \over 2}(1-{1\over \sqrt{3}}) \simeq 0.2113  .\nonumber
\ena

Two remarkable properties of Gaussian random fields emerge 
from the above analytical calculations: 
\bi
\item the evolution of the ``hill'', ``lake'' and ``saddle'' point
fractions as functions of threshold $\delta_{\rm th}$ depends 
only on the spectral parameter
$\gamma$; 
\item the asymptotic value for $\delta_{\rm th}\rightarrow
-\infty$ is {\em independent} of any spectral parameter, \ie {\em is the same for
any Gaussian random field}. This would not be true if we were
considering only maxima, minima or saddle points (see BE87).
\ei

\begin{figure}
\centering
\centerline{\epsfysize=\hsize \epsfxsize=\hsize\epsfbox{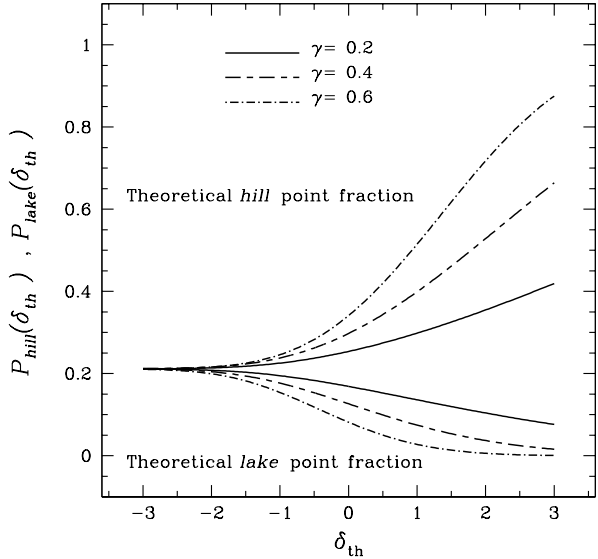}}
\caption{Functions ${\mathcal P}_{\rm hill}(\delta_{\rm th})$ and
${\mathcal P}_{\rm lake}(\delta_{\rm th})$ in the Gaussian case for various values of
$\gamma$ as indicated on the figure. Their value is independent of $\gamma$ in the limit 
$\delta_{\rm th}\rightarrow -\infty$, but their
evolution with $\delta_{\rm th}$ is rather sensitive to $\gamma$.
Similar conclusions hold for function ${\mathcal P}_{\rm saddle}(\delta_{\rm th})=
1-{\mathcal P}_{\rm hill}(\delta_{\rm th})-{\mathcal P}_{\rm
lake}(\delta_{\rm th})$, which is not plotted for simplicity.}  
\label{plot_th_frac_gamma_1}
\end{figure}

To illustrate these results, we draw on Fig.~\ref{plot_th_frac_gamma_1} the functions ${\mathcal
P}_{\rm lake}(\delta_{\rm th})$ and ${\mathcal P}_{\rm hill}(\delta_{\rm th})$ for a set of $\gamma$
values. Except in the limit $\delta_{\rm th}\rightarrow -\infty$, the
evolution of these functions with $\delta_{\rm th}$ is rather sensitive to $\gamma$. 

At this point, it is important to be aware that till now, we
have supposed an idealistic, infinite resolution experiment, \ie we did not take
into account any \emph{beam smearing} effect that would affect any
real measurement. Consequently, in order to compare the predictions to any true
measurements, it would be necessary to incorporate this effect in our
calculations. To do so, we should consider the convolved field, $\delta \ast B$,
where $B$ stands for the instrumental beam response, instead of the
idealistic fluctuation field, $\delta$. Since linear
transformations such as the convolution by an arbitrary beam function
do not change the Gaussian nature of a random field,
the beam smearing should appear only as a dependence of $\gamma$ on the
smoothing scale $\sigma_{\rm b}$, that we will note
$\gamma(\sigma_{\rm b})$. Furthermore, although the problem of beam 
convolution in its full generality is very intricate, it turns out to be analytically
tractable for a Gaussian, symmetric beam, a reasonable approximation
in practice. We will illustrate this point in more details below.

\section{Confronting Predictions and Measurements on Simulations}
\label{measure}

In order to test our ability to measure the functions ${\mathcal P}_{\rm
hill}(\delta_{\rm th})$ and ${\mathcal P}_{\rm lake}(\delta_{\rm
th})$,\footnote{Again, we restrict here our analysis to these two functions,
since function ${\mathcal P}_{\rm
saddle}(\delta_{\rm th})$, which is equal (by construction) to 
$1-{\mathcal P}_{\rm hill}(\delta_{\rm th})-{\mathcal P}_{\rm
lake}(\delta_{\rm th})$, does not yield any further information.} and
to measure the incertainties for maps of limited extent,
we performed several measurements on simulated maps. We henceforth limit ourselves to small 
simulated square patches of the sky of width $\sim 12.5\
\mathrm{deg}$. These patches are considered as being flat and
pixelized with a $512 \times 512$ Cartesian mesh of resolution
$\theta_{\rm pix} = 1.5^{\prime}$.  

This section is organized as follows: we first describe the measurement principles ({\S}~\ref{pple_meas})
and apply them to noise free realization of Cold Dark Matter (CDM)
($\Omega_m = 0.3$, $\Omega_{\Lambda} = 0.7$, $h=0.5$, $\Omega_b =
0.05$ and $n=1$) Gaussian maps generated from their power spectrum (taking into account
both the uniform distribution of the phases of the Fourier modes and
the Rayleigh distribution of their modulus) \footnote{The power
spectrum has been computed making use of the publicly available
CMBFAST code \cite{SeZa96} available at {\texttt
http://physics.nyu.edu/matiasz/ CMBFAST/cmbfast.html}}
({\S}~\ref{noifree_app}); in particular, we examine the Gaussianity of
the distribution function of measurements in order to be able to
perform later $\chi^2$ tests; finally we discuss practical issues
related  to finite resolution and finite volume effects
({\S}~\ref{gamma_eff}). 

\begin{figure}
\centering
\centerline{\epsfysize=\hsize \epsfxsize=\hsize \epsfbox{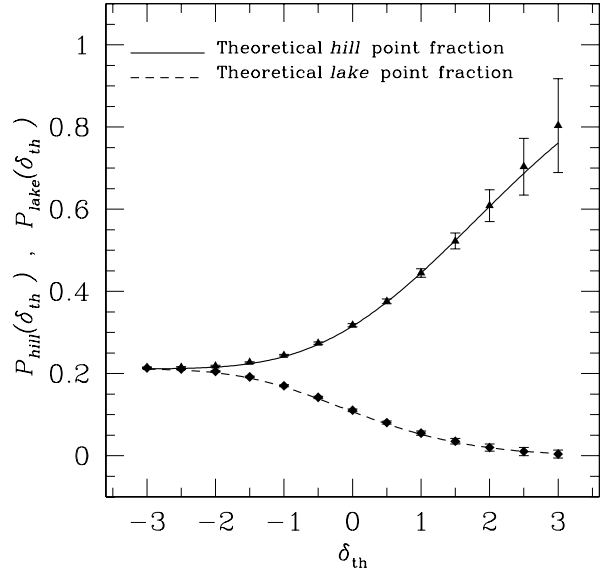}}
\caption{Measurement of functions ${\mathcal P}_{\rm hill}$ and
${\mathcal P}_{\rm lake}$ on Gaussian simulated maps in the standard CDM case. 
The triangles and the losanges give respectively the hill and lake point fractions
obtained from an average over 200 realizations of size $12.5\times 12.5\
\mathrm{deg}^2$. The error bars on each symbol are obtained from the 
dispersion over the 200 realizations. Prior to measurement, a
smoothing with a Gaussian window of size  $\sigma_{\rm b} = 5\theta_{\rm
pix}$ was performed. The continuous and dashed curves 
correspond to the theoretical expectations for a Gaussian random field with $\gamma_{\rm eff} = 0.48$.
\label{plot_th_frac_err_1}}
\end{figure}

\subsection{Principles of measurements}
\label{pple_meas}
The measurements consist in determining the fractions
${\mathcal P}_{\rm hill}$ and ${\mathcal P}_{\rm lake}$  
for an ensemble of excursion sets (subsets for which 
$\delta \ge \delta_{\rm th}$) of an image smoothed
at different scales $\sigma_{\rm b}$. To insure sufficient
differentiability, i.e. at least continuity of second order
derivatives, we choose a Gaussian smoothing window.
 
Similarly as in Colombi et al. (2000) for the 3D case,
to measure the local curvature at a given point, we find the quadratic
form which fits this point and its $8$ closest neighbors. 
We deduce from the quadratic form coefficients the local values of the second
derivatives, \ie the Hessian matrix, then diagonalize this matrix and
determine the sign of its eigenvalues $\lambda_1$ and $\lambda_2$. Both easy
to implement and fast, this method turns out to be quite robust in
the presence of noise.

\subsection{Example: application to noise free realizations}
\label{noifree_app}

As a first example, we apply this measurement technique to $200$
simulated noise free realizations of CDM like temperature maps
smoothed by a Gaussian beam of $\sigma_{\rm b} = 5\theta_{\rm pix} =
7.5^{\prime}$. In each realization, we measure both ${\mathcal P}_{\rm
hill}(\delta_{\rm th})$ and ${\mathcal P}_{\rm lake}(\delta_{\rm th})$
for the ensemble of excursion sets defined by $\delta_{\rm th}= -3.0$,
$-2.5$, $-2.0\ldots 3.0$, as illustrated by Fig.~\ref{plot_th_frac_err_1}. For comparison purpose,
we plotted on this figure the theoretical curves corresponding to the
{\em expected} value of the parameter $\gamma (\sigma_{\rm b} =
5^{\prime}) = 0.48$. This effective $\gamma$ can be easily calculated for a Gaussian beam, 
since in this context we just have to replace $P(k)$ (an exact
theoretical input here) by $P(k)\exp(-k^2\sigma_{\rm b}^2)$ in
Eqs.~(\ref{sig2j}) and (\ref{gammadef}),  (see Fig.~\ref{plot_gamma_effvsmes_smooth} and section
{\S}~\ref{gamma_eff} for a mode detailled illustration of this
calculation). With this choice of $\gamma$, the theoretical curves fit
very well the data. The dispersion over the measurements is very tight
except at high threshold where we enter a rare events regime, as
indicated by the larger $1 \sigma$ error bars. 
 
To illustrate more visually the protocol, Fig.~\ref{panel_cdm} displays
one CDM realization (noise free again) smoothed at $\sigma_{\rm b}= 3\theta_{\rm pix}$, 
the corresponding excursion set for $\delta \ge 1$ as well as the 
local curvature map, where $hill$, $lake$ and $saddle$ points are
identified by respectively black, white and grey points. A 
cross examination of these images provides some insight on the
processe involved, \ie the correlation of extrema and local curvature.  

\begin{figure}
\centering
\centerline{\epsfxsize=0.5\hsize \epsfbox{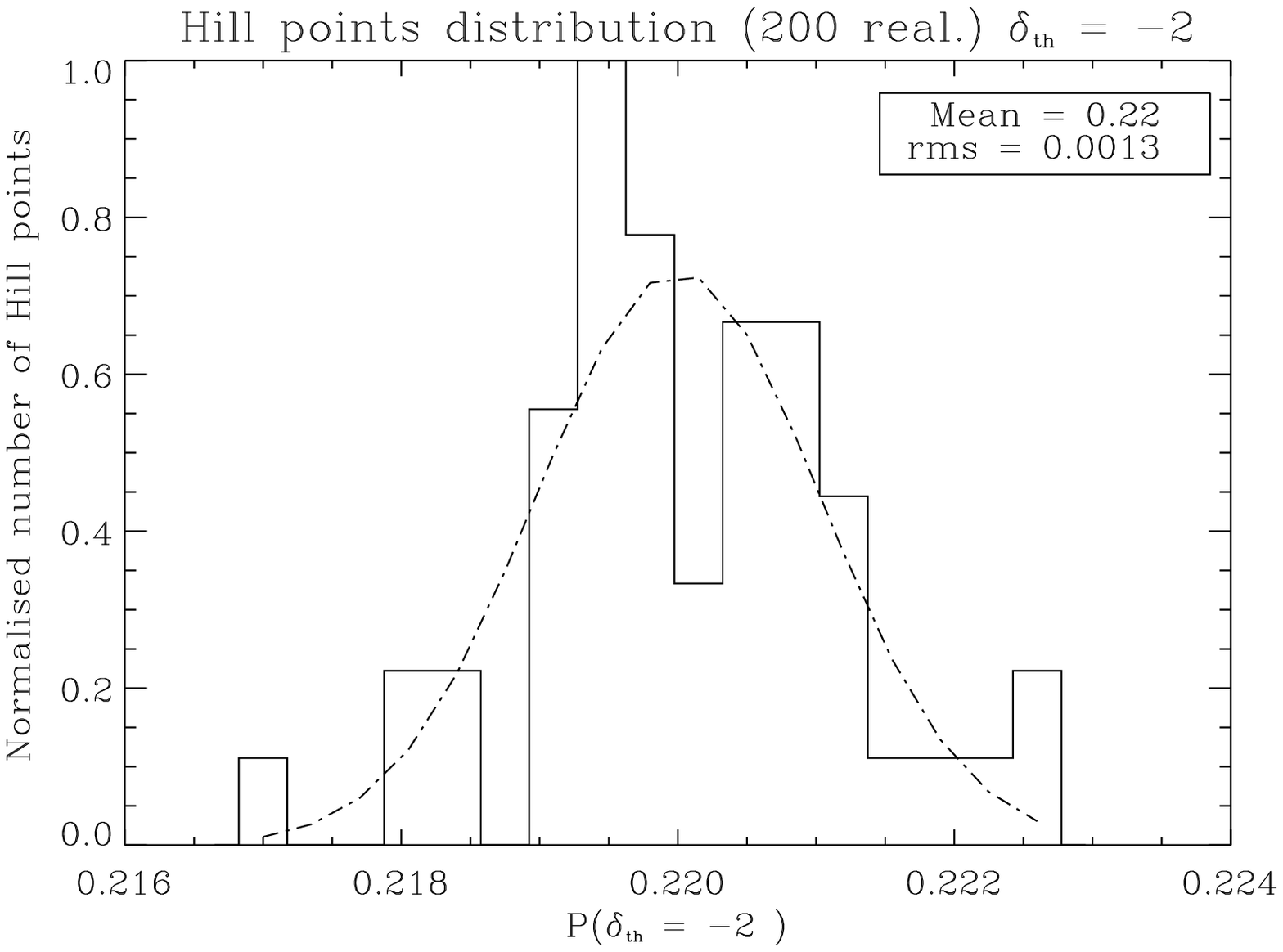}\epsfxsize=0.5\hsize
\epsfbox{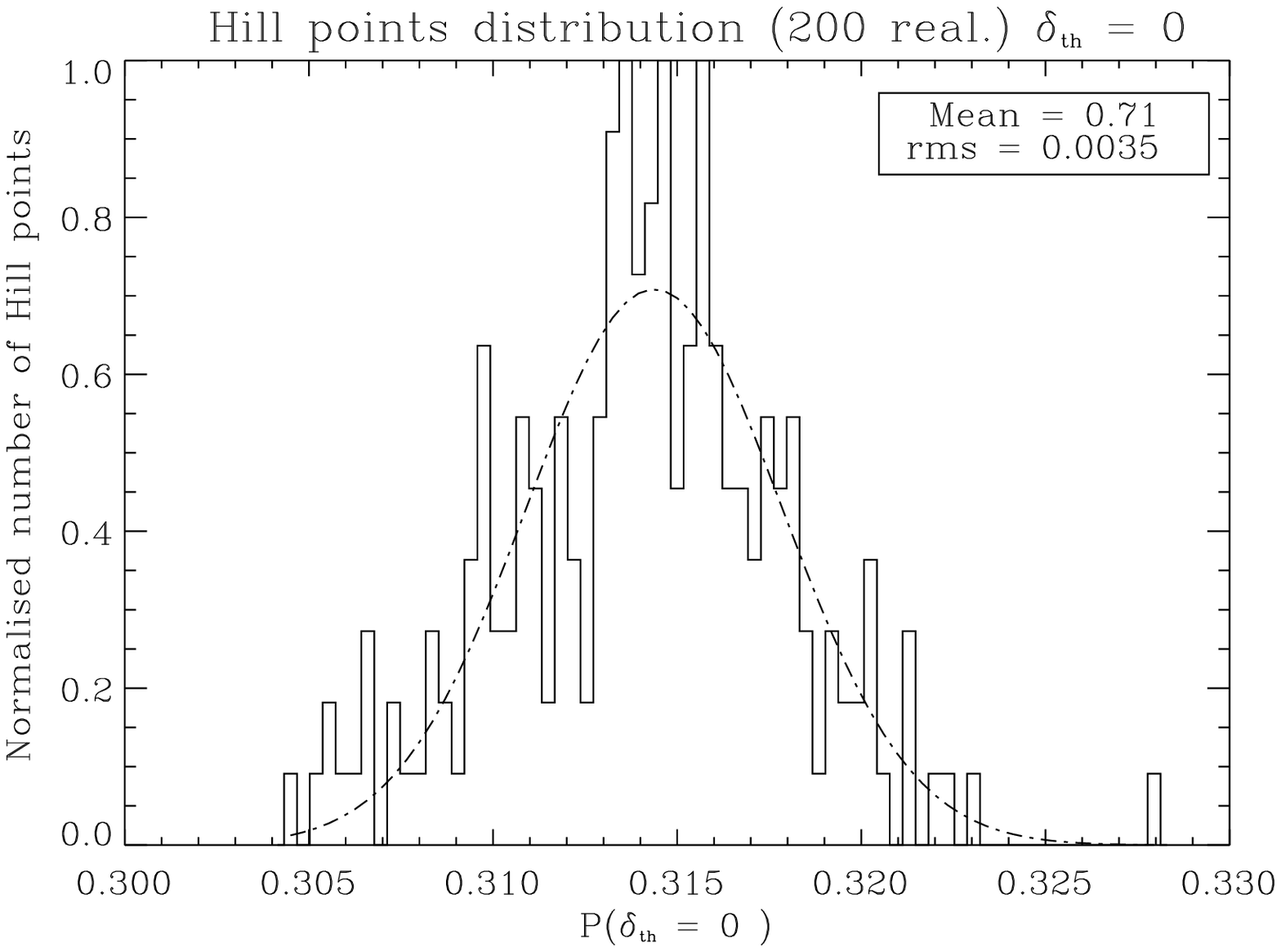}}
\caption{Distribution function of hill points fraction obtained from
200 CDM noise free realizations of temperature maps smoothed with a
gaussian window of size $\sigma_{\rm b} = 3\theta_{\rm pix}$,
similarly as in Fig.~\ref{panel_cdm}. Two excursion sets are
considered, one with $\delta_{\rm th} = -2$ (left panel) and the other
one with $\delta_{\rm th} = 0$ (right panel). The smooth dotted-dashed 
curve on each panel corresponds to a Gaussian of same variance as the
distribution function. It superposes well to the measurements
(histograms) for the excursion sets considered here.\label{distrib_hillp} }
\end{figure}

To further investigate the dispersion over the measured values of 
${\mathcal P}_{\rm hill}(\delta_{\rm th})$ at a given $\delta_{\rm
th}$, we drawn on Fig.~\ref{distrib_hillp} the probability distribution function of the hill point
fraction measured from the 200 realizations [similar results can
be obtained for ${\mathcal P}_{\rm lake}(\delta_{\rm th})$]. Two thresholds are
considered,  $\delta_{\rm th} = -2$ (left pannel) and $\delta_{\rm
th}=0$ (right pannel). The dashed line on each pannel corresponds to
the Gaussian limit with same standard deviation as the one actually
measured. It fits qualitatively well the measurements, so the
previously drawn $1\,\sigma$ errors are meaningful estimates of the
cosmic variance errors.\footnote{However, one must keep in mind that error bars
for different values of $\delta_{\rm th}$ are correlated with each
other, but this will be taken into account in the analyses conducted
later.} We checked wether the distribution function of the measurements is
Gaussian for other values of $\delta_{\rm th}$ and found that it is
indeed the case except at high threshold, $\delta_{\rm th} \ge 2.5$,
where one enters the rare events regime.  

Thus, the measurements in this \emph{idealistic} noise free case are in very good agreement with
theoretical expectations. Furthermore, the cosmic errors associated
with these measurements are very tight for a Gaussian field. This last point will be of
particular interest to us since it makes the evolution of functions
${\mathcal P}_{\rm hill}(\delta_{\rm th})$ and ${\mathcal P}_{\rm
lake}(\delta_{\rm th})$ with $\delta_{\rm th}$ a sharp signature of random Gaussian fields.

\setcounter{figure}{4}
\begin{figure*}
\centering{
\centerline{\epsfxsize=11cm \epsfbox{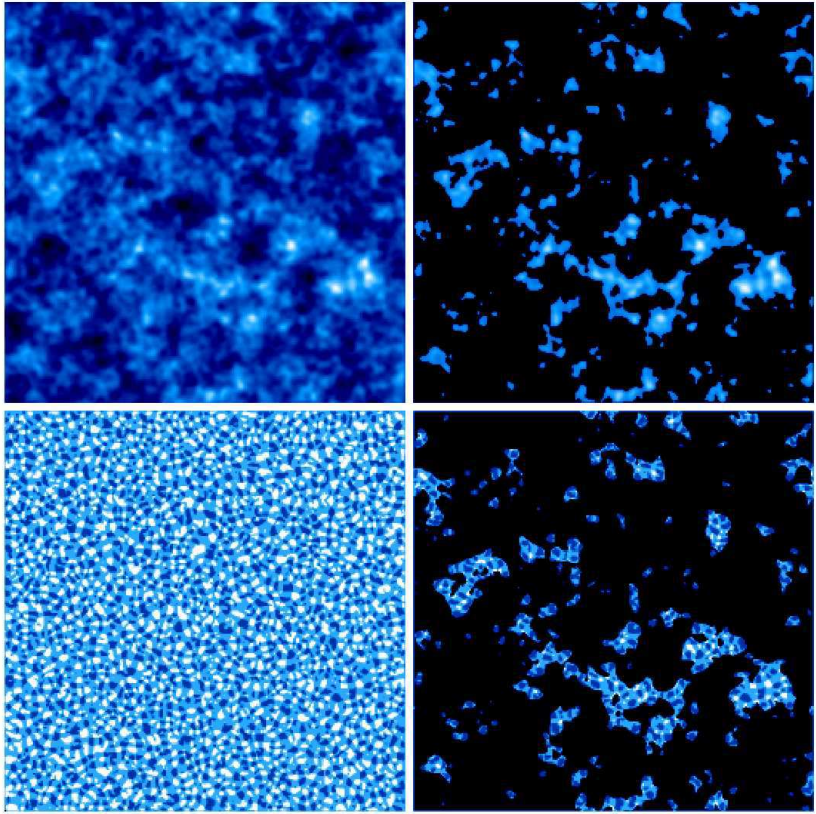}}
\caption{Illustration of the measurement protocol. Considering a CDM 
temperature map with $512 \times 512$ pixels, 
of width $12.5\ \mathrm{deg}\times 12.5\ \mathrm{deg}$ and
smoothed with a Gaussian window of size 3 pixels (top left panel),
we identify ``hill'', ``lake''and ``saddle'' points according
to their local curvature (respectively by dark grey, 
white and light grey points on the
bottom left panel). The fraction of space occupied by each 
kind of point is studied as a 
function of the density threshold. As an example, the excursion set
corresponding to $\delta_{\rm th} = 1$ is shown in right pannels,
which are the same than left panels except that only regions with
$\delta \ge \delta_{\rm th}$ are shown, the rest being coded
in black. As one can see, the relative abundance of ``lake'' 
points in bottom right panel is now smaller than
on bottom left panel, while which of ``hill'' points has augmented, as expected.
The measurement of the relative fraction of the three kind of points as a function
of density threshold is the key idea of this paper.
\label{panel_cdm}} }
\end{figure*}

\subsection{Measuring $\gamma$: spurious effects and available dynamic range}
\label{gamma_eff}

In the above discussion, we used a \emph{predicted} value of $\gamma$ to check
if the measured ``hill'', ``saddle'' and ``lake'' point fractions 
agreed with theoretical predictions for a Gaussian random field. 
Therefore, the only thing we proved so far is that for a finite realization of a
Gaussian random field smoothed with a Gaussian of width $\sigma_{\rm b} = 3\theta_{\rm
pix}$, whose (unsmoothed) power spectrum is perfectly known, the
predicted value $\gamma$ agrees very well with the measured points. In
practice however, the knowledge of the power spectrum might be less
accurate and we want to develop a non-Gaussianity test which  
can be performed both independently or in combination with the power
spectrum measurement. Thus it is important to determine from the
measurements of these fractions themselves an effective value,
$\gamma_{\rm eff}$, that fits well the data. We will see that for a
finite realization of a Gaussian random field, there always exists a value
$\gamma_{\rm eff}$ which is such that measurements agree very well
with analytic predictions.  

As we shall see below, the determination of the $\gamma_{\rm eff}$
value is easy and this fact, by itself, is highly 
significant and is the key point of our paper: for a map in which temperature
fluctuations are partly seeded by cosmic strings, we shall see that it
is actually not possible to find a value of $\gamma_{\rm eff}$ leading
to point fractions matching the measurements in all  the available
dynamic range -- namely, for all possible values of $\delta_{\rm th}$ 
and $\sigma_{\rm b}$.  However, as we shall see later, it is not needed
to accurately  determine the real value of $\gamma$ to 
efficiently constrain the level of non Gaussianity of a map.

However, studying the difference between $\gamma$ and $\gamma_{\rm
eff}$ can help to determine the available dynamic range,
i.e. the set of values of $\delta_{\rm th}$ and $\sigma_{\rm b}$ 
in which one can trust the measurements. We already noticed in previous section that
the density threshold should be small enough, $\delta_{\rm th} \la
2.5$, to avoid entering the rare event regime, where the cosmic
distribution function of measurements becomes non Gaussian.
Here, we are concerned by two effects that we ignored
previously: 
\begin{itemize}
\item Pixelization effects, or equivalently, effects of finite
resolution: as discussed above, it is important to smooth the data
to insure sufficient differentiability. The smoothing scale should
be large enough compared to the pixel size to avoid
anisotropies or discreteness effects brought by the pixelization.
\item Finite volume and edge effects: our experimental maps cover a
rather small fraction of the sky, due to the limited dynamical range
in the cosmic-string simulations from \cite{BoBe88,BeBo90}. Therefore,
to avoid reducing too much the number of statistically independent
regions of the map, the smoothing scale should not be too large. In a
more realistic experiment covering a large fraction of the sky, 
finite volume and edge effects should not be as much as of a concern.
\end{itemize}

The practical measurement of $\gamma_{\rm eff}$ is made straightfoward by
noticing that the dependence of the function $\tilde{\mathcal P}_{\rm
hill}(\delta_{\rm th}=1)$ on $\gamma$ is very well approximated by an
exponential law\footnote{Note that such a fit is also appropriate for values of
$\delta_{\rm th}$ different from 1.} in the domain of interest to us, \ie $0.4 \le \gamma \le 0.95$: the
fit  
\be
\ln \tilde{\mathcal P}_{\rm hill}(\delta_{\rm th}= 1) \simeq a + b\gamma
\en
with $a= -1.4099$ and $b=1.2395$ is accurate to $0.6\%$.
The choice of the particular value  $\delta_{\rm th}= 1$ is an ad hoc
compromise coming from the competition between two effects: (i) the
dependence on $\gamma$ of the quantity $\tilde{\mathcal P}_{\rm
hill}(\delta_{\rm th})$ increases with $\delta_{\rm th}$
(Fig.~\ref{plot_th_frac_gamma_1}), but (ii) the uncertainty on the
determination of $\tilde{\mathcal P}_{\rm hill}(\delta_{\rm th})$
increases with $\delta_{\rm th}$ (Fig.~\ref{plot_th_frac_err_1}). 

To test the effects of finite coverage and finite resolution, 
we examine a CDM case, where we generated once again some Gaussian field
realisations from their $C_{\ell}$. Assuming as before that the field has been
smoothed by a Gaussian of width $\sigma_{\rm b}$,  we can
easily compute the prediction for the function $\gamma(\sigma_{\rm b})$.

\begin{figure}
\centering
\centerline{\epsfxsize=\hsize {\epsfbox{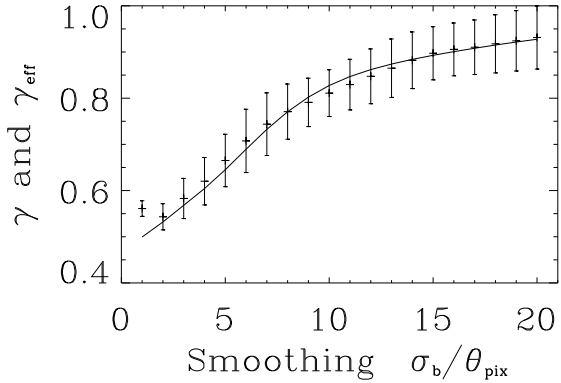}}}
\caption{Measurement of $\gamma_{\rm eff}$ as a function of the
smoothing scale $\sigma_{\rm b}/\theta_{\rm pix}$. The solid line gives the theoretical
values of $\gamma$ for various $\sigma_{\rm b}$. They were computed
using the power spectrum $C_{\ell}$. The crosses with error bars correspond to
measurements of $\gamma_{\rm eff}$ using the measurements of
${\cal P}_{\rm hill}(\delta_{\rm th}=1.0)$. The error bars correpond
to $\pm 1$ the measured rms on 50 realisations. Pixelisation effects bias
the measurements of $\gamma_{\rm eff}$ till $\sigma_{\rm
b}/\theta_{\rm pix} = 3 $. The error bars grows with $\sigma_{\rm b}$ as expected from
sampling variance arguments. 
\label{plot_gamma_effvsmes_smooth}}   
\end{figure}

In Fig.~\ref{plot_gamma_effvsmes_smooth}, the measured values of $\gamma_{\rm
eff}$ is displayed in each case as a function of smoothing
scale. An average over 50 realizations of $512\times 512$ pixels maps
is performed. The error bars on the figure represent the corresponding
scatter, which increases with $\sigma_{\rm b}$ as expected, due to
finite volume effects. The solid line corresponds to the theoretical
function $\gamma(\sigma_{\rm b})$. Agreement between $\gamma_{\rm
eff}$ and the analytic prediction is good, except at small scales
where pixelisation effects contaminate this  measurements.
We however see that pixelisation effects become negligible when $\sigma_{\rm b}/\theta_{\rm pix} >
3$. Note that there should be an upper bound as  well for $\sigma_{\rm
b}$ as discussed above, since cosmic errors 
increase with scale. Our $\chi^2$ analysis below will anyway naturally
take that into account  by giving a lesser weight to scales with larger
errors. This reasonable agreement between the measured $\gamma_{\rm
eff}$ at a given scale $\sigma_{\rm b}$ and the expected
$\gamma(\sigma_{\rm b})$ validates the approach we will use in the
next section which consists in finding from the fraction themselves an
ad hoc $\gamma_{\rm eff}$, and deducing from it an agreement with Gaussianity.

\section{Testing Non-Gaussianity : Principle and Application to Mixed Models}
\label{ngtest}

>From the previous section, we conclude that in the case of a 
smoothed Gaussian random field, the functions ${\cal P}_{\rm hill}(\delta_{\rm
th})$ and ${\cal P}_{\rm void}(\delta_{\rm th})$ can be measured accurately and fit 
very well the analytical predictions provided $\gamma$ be
considered as an adjustable parameter. Furthermore, their 
probability distribution function is well approximated by a Gaussian if
$\delta_{\rm th} \la 2.5$, which now allows us to define a rigorous
measurement protocol based on $\chi^2$ analysis that we shall apply
to simulated data in the trustable scale range determined above, 
namely $\sigma_{\rm b} \ge 3$.

In this section, we first detail how we use the functions  ${\cal P}_{\rm hill}(\delta_{\rm
th})$  and ${\cal P}_{\rm void}(\delta_{\rm th})$ 
to build a $\chi^2$ statistic testing
Gaussianity ({\S}~\ref{ng_method}). Then we apply the method to 
simple simulated mixed models
involving cosmic strings plus CDM and see how it can be used to bound
quantitatively the relative contribution of cosmic strings ({\S}~\ref{appmix}). 

\subsection{The method}
\label{ng_method}

Let us assume as before that we have at our disposal an observed temperature
map $\delta_{\rm T}$, on which we measured the fractions
${\mathcal P}_{\rm hill}(\delta_{\rm th_i},\sigma_{\rm b_i})$ and
${\mathcal P}_{\rm lake}(\delta_{\rm th_i},\sigma_{\rm b_i})$ for a set of
threshold values  $\{\delta_{\rm th_i}\}$ and smoothing scales $\{\sigma_{\rm b_i}\}$.
If the power spectrum of this random field is known, we can in
principle define analytically a set of values of $\gamma$ for each smoothing scale,
$\{\gamma_{\sigma_{\rm b_i}}\}$. Assuming furthermore that the field is
Gaussian and that our estimators of ${\mathcal P}_{\rm hill}$ and
${\mathcal P}_{\rm lake}$ are unbiased, it is
easy to deduce from this $\gamma$ set some \emph{theoretical
expectations} for the set, 
$\tilde{\mathcal P}_{\rm X}(\delta_{\rm th_i},\sigma_{\rm b_i}) \equiv 
\langle {\mathcal P}_{\rm X}(\delta_{\rm th_i},\sigma_{\rm b_i}) \rangle$,
where ${\rm X}$ stands from now on for ``lake'' or ``hill''. To quantify the distance
between these theoretical predictions and the measurements, we
introduce the standard $\chi^2$ statistic. Since
these measurements can obviously not be considered as independent, we
have to introduce the full theoretical variance-covariance matrix
\begin{equation}
C_{\rm II'} \equiv \langle({\mathcal P}_{\rm I}-
\tilde{\mathcal P}_{\rm I}) 
({\mathcal P}_{\rm I'} 
-\tilde{\mathcal P}_{\rm I'}) \rangle,
\label{c_mat}
\end{equation}
where we use the short hand notation ${\cal P}_{\rm I}=
{\mathcal P}_{\rm X}(\delta_{\rm th_i},\sigma_{\rm b_i})$ and ${\cal P}_{\rm I'}=
{\mathcal P}_{\rm X'}(\delta_{\rm th_{i'}},\sigma_{{\rm B_{i'}}})$.
Then, the statistic 
\be 
\chi^2 \equiv \sum_{\rm I I'}\ ({\mathcal P}_{\rm I}- \tilde{{\mathcal P}}_{\rm I})\
(C_{\rm II'})^{-1}\ ({\mathcal P}_{\rm I'}- \tilde{{\mathcal P}}_{\rm I'})\ . 
\label{chi2}
\en
is expected to follow a $\chi^2$-distribution, 
as illustrated by a practical example below. 
Indeed results of {\S}~\ref{noifree_app} suggest that the
distribution function of the measured 
${\mathcal P}_{\rm I}$ is nearly 
Gaussian if $\delta_{\rm th} \la 2.5$.

In practice, we have to introduce two more subtleties in this
protocol. 

First, as discussed extensively in {\S}~\ref{gamma_eff}, the practical
realization of a Gaussian random field yields a measured function 
${\mathcal P}_{\rm I}$ matching the theory, $\tilde{\mathcal P}$, but with 
an {\em effective} value of $\gamma$. Taking into account some
additional Gaussian noise, as discussed in more details below, would
make this effect even stronger. Rigorously, we should modelize this
$\gamma_{\rm eff}$ effect in terms of statistical bias on the
estimators of the function ${\mathcal P}_{\rm X}(\delta_{\rm th},\sigma_{\rm b})$, but we proceed here
differently, for simplicity: we extract the $\gamma_{\rm eff}(\sigma_{\rm b})$
from the data by fitting the analytic prediction for function 
${\mathcal P}_{\rm hill}(\delta_{\rm th}=1,\sigma_{\rm b})$ 
to the measured one. Thus, our ``theory'' depends itself on
the data through $\gamma_{\rm eff}$. As already explained,
the critical test for non Gaussianity will be on the evolution
of the functions ${\mathcal P}_{\rm X}(\delta_{\rm th},\sigma_{\rm b})$ with $\delta_{\rm th}$,
which, once $\gamma_{\rm eff}$ is determined, is entirely fixed. 
The practical measurement of $\gamma_{\rm eff}$ is performed as
explained in the previous section.  In principle we could include the
value of $\delta_{\rm th}$ used to measure $\gamma_{\rm eff}$  
as a varying parameter in our $\chi^2$ test, to optimize the
analysis. Given the level of accuracy of our numerical experiments and
since we just aim here to illustrate the method in a simple and
convincing way, we did not feel necessary to do so. 

Second, even if in principle we could compute analytically or numerically the
variance-covariance matrix, for the sake of simplicity we evaluated
it using a Monte-Carlo method based on a few hundreds of
{\em Gaussian} realizations having the same power-spectrum
and noise properties as the input map. So for each model we will
consider below (pure CDM, mixed model with CDM + cosmic strings,
and pure cosmic strings), the covariance matrix $C$ should be different
in each case. However, in practice we checked that this matrix coefficients depend very slightly on $\gamma_{\rm eff}$ in the range of interest to us, \ie $0.4 \le \gamma \le 0.95$ so that we consider for simplicity only the same $C$ matrix in each case.
Another issue concerning this matrix is that it might be singular. Indeed, since
fractions measured at various thresholds and different scales can be 
highly correlated, \eg the values at very low threshold where the
dependence on $\gamma$ is very weak, the studied $\{\delta_{\rm th_i}\}$
and $\{\sigma_{\rm b_i}\}$ sets have to be restricted so that $C$ is invertible.

\subsection{Application on mixed models for MAP and Planck}
\label{appmix}

As an illustration of the accuracy of our statistics, we apply it in the
framework of MAP\footnote{\texttt{http://map.gsfc.nasa.gov}} 
and Planck Surveyor\footnote{\texttt{http://astro.estec.esa.nl/SA-general/Projects/Planck/}}
experiments to a mixed model where temperature fluctations are seeded in part by
cosmic strings and otherwise by adiabatic inflationary perturbations
in standard CDM model (\eg Bouchet \etal 2001 and references
therein). Our approach is rather simple, since we add together 
a CDM map and a temperature fluctuation map obained from 
ray-tracing in cosmic string simulations \cite{BoBe88,BeBo90}
and neglect possible cross-correlations between such maps.
This implicitely supposes the existence of a standard CDM
inflationary epoch followed by a phase transition during which cosmic
strings appear and then imprint supplementary fluctuations 
on the dark matter distribution. Neglecting cross-correlations
mentionned just above is equivalent to assume that the
distribution of cosmic string as well as their dynamical evolution
is not coupled with that of cold dark matter, which should be a good
assumption at the level of approximation considered in this paper
\cite{LiRi97,CoHi98}. We neglect also any contribution to the
fluctuations prior to the last scattering surface (lss) so that the
comparaison with results from Bouchet \etal 2001, who considered this
contribution, is not immediate at small scales. 

To take into account the noise expected in the best channels of 
MAP and Planck Surveyor experiments
(respectively $12.8\ \mathrm{\mu K/K}$ per $0.3\times 0.3\ \mathrm{deg}^2$ pixels and $2\
\mathrm{\mu K/K}$ per $8'\times 8'\ \mathrm{deg}^2$), we add an extra Gaussian
white noise $n$ to our square maps.  Note again, that as long as the noise is
Gaussian, its presence can be simply taken into account by 
a change in the value of $\gamma_{\rm eff}$.  Thus, our simulated temperature maps read
\be
\delta_{\rm T} = (1-\beta)\ \delta_{\rm T}^{\rm string} + \beta\ \delta_{\rm T}^{\rm CDM} + n,
\en
where $1-\beta$ represents the fraction of the signal seeded by cosmic strings, given
the fact that we impose ${\rm rms}(\delta_{\rm T}^{string}\ ) =
{\rm rms}(\delta_{\rm T}^{CDM}) = {\rm rms}(\delta_{\rm T}^{COBE}\ )$
for the fields smoothed at a scale corresponding to the COBE-DMR beam,
\ie $\sigma_{\rm b}=7^\circ/\sqrt{8\ln 2}$. Note that we differ from
Bouchet \etal 2001  in the relative normalisation of the two
contributions since they define their relative normalisation with
$C_{\ell}$, \ie they considered a family of models parametrized by
$C_{\ell} = (1-\alpha)\ C_{\ell}^{\rm string} + \alpha\ C_{\ell}^{\rm
CDM}\displaystyle $, where both $C_{\ell}^{\rm string}$ and
$C_{\ell}^{\rm CDM}$ are COBE normalised and where $C_{\ell}^{\rm
string}$ includes contributions from before and after the lss. If we
wanted however to express our results, in terms of $\beta$, we should consider $1-\alpha =
(1-\beta)^2/((1-\beta)^2 + \beta^2)$.   

We now construct the previously defined $\chi^2$ statistics for various
values of $\beta$ and try to figure out what is the smallest
``string like contribution'' that our method should be able to detect
with a high statistical significance. 

\subsubsection{Constructing the $\chi^2$ and adequacy of the $\chi^2$ probability distribution}
\begin{figure}
\centering
\centerline{ \epsfxsize=\hsize \epsfbox{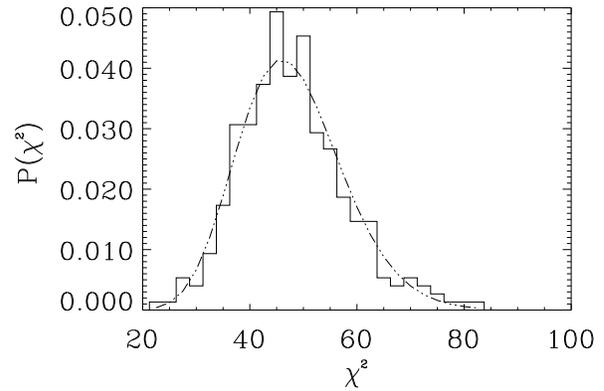}}
\caption{Histogram of the measured probability distribution of the
quantity $\chi^2$ defined by Eq.~(\ref{chi2}), from $300$ noisy (Planck best
channel noise level) realizations of CDM temperature fluctuations. 
The dot-dashed curve corresponds to a $\chi^2$-distribution with $47$ degrees of
freedom (dashed line). The good agreement allows us to use 
$\chi^2$ as a measure of the Gaussian nature of the observed
field.\label{plot_chi2_pdf_nop_cdm}}  
\end{figure} 

Even if one would like to include as many threshold values and smoothing
scales as possible in the analysis, this is in practice not
necessary. Indeed, some measurements are, at least in the Gaussian case, 
highly correlated with each other, due to \eg some strong
constraints (in the very low threshold regime,
$\delta_{\rm th}\rightarrow -\infty$, 
both ${\mathcal P}_{\rm lake}(\delta_{\rm th})$ and
${\mathcal P}_{\rm hill}(\delta_{\rm th})$ tend to an
asymptotic value independent of the spectral parameters).
Therefore, one can restrict the set of values of
$(\sigma_{B},\delta_{\rm th})$ by requiring that
the $C$ matrix (see Eq.\ref{c_mat}) be non-singular. 
 
By investigating the numerical properties of $C$, 
we find for all the maps \footnote{As already stated, we checked that
this choice is adequate for various $C$ matrices, \ie corresponding to
different ${\gamma}$s, altough we use only one in practice.} that
$\delta_{\rm th} = -2$,$-1.5\ldots$, $+1.5$ and  
$\sigma_{\rm b}/\theta_{\rm pix} = 6$, $10$ and $14$ (corresponding to
repectively $21'$, $35'$ and $49'$ on the sky) are satisfying values for
the Planck case, whereas $\delta_{\rm th} = -1.5$,$-1.0\ldots$, $+1.5$ and
$\sigma_{\rm b}/\theta_{\rm pix} = 8$ and $12$ (corresponding to
respectively $28'$ and $42'$) are reasonable for MAP. Of
course, this choice of the set of values of $(\sigma_{B},\delta_{\rm th})$
is strongly influenced by the fact
that our map is rather small: for a full sky survey, the sampling
would probably be different. 

Given these restrictions, it is now important to explore the properties of
the probability distribution function of the $\chi^2$ statistics defined in
Eq.~(\ref{chi2}). As stated previously, we expect it to follow a
$\chi^2$-distribution, at least in the Gaussian case, 
since the distribution of measured values of ${\mathcal P}_I$s is well
described by a Gaussian (see {\S}~\ref{pple_meas}). 
We tested this by generating 300 realizations 
of noisy (Planck level) CDM like only temperature fluctuations
for which we measure $\chi^2$. Its probability distribution is
drawn on Fig.~\ref{plot_chi2_pdf_nop_cdm}, where we superimposed a
theoretical $\chi^2$-distribution with $47$ degrees of freedom (dof) ($48$
measurements: $6\times 4$ measured hill 
points fraction and $6\times 4$ measured void point fractions;  
$1$ parameter, $\gamma$). The agreement is obviously
very good, comforting the analysis procedure and making this
$\chi^2$ statistic adequate for assessing Non-Gaussianity.\footnote{Here,
we do not test if the $\chi^2$
estimator always follows a $\chi^2$-distribution, \ie 
also for non-Gaussian maps. Such a test is not necessary since the prior
we use to analyse the data assumes underlying Gaussianity (it would
be impossible to do otherwise in practice).}

\subsubsection{Ability to distinguish a ``Non-Gaussianity Level'' with
a realistic noise level}

\begin{figure}
\centering
\centerline{\epsfxsize=1.15\hsize \epsfbox{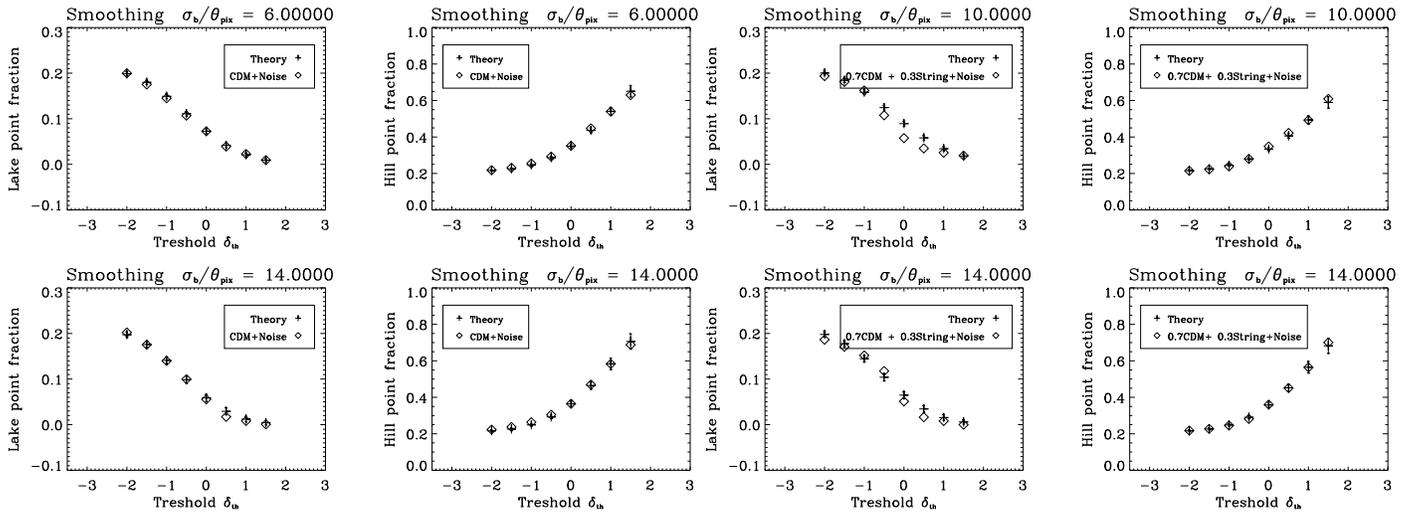}}
\caption{Measurement of hill and lake point fractions in a Planck simulation with pure CDM 
($\beta=1.0$). The left column shows the {\it measured} evolution of ${\mathcal
P}_{\rm lake}$ as a function of threshold, $\delta_{\rm th}$, for two smoothing
scales: top left panel corresponds to $\sigma_{\rm b}/\theta_{\rm pix} = 10$
and bottom left panel to $\sigma_{\rm b}/\theta_{\rm pix} =
14$. Symbols without error bars represent the measurements in the
realization, while symbols with errorbars correspond to the Gaussian
prediction matching  the value of ${\mathcal P}_{\rm hill}(\delta_{\rm
th}=1)$ as explained more in detail in {\S}~\ref{ng_method}. 
The error bars are obtained from the 1-$\sigma$ dispersion over 300
realizations.  The right column of panels is similar to left column, but for
 ${\mathcal P}_{\rm hill}$. \label{plot_cdmt_topo2_cdm_noP}}
\end{figure}

Making use of this $\chi^2$ statistics,  we are now in position to determine how well 
we can distinguish non-Gaussian signatures. 

\begin{figure}
\centering
\centerline{\epsfxsize=1.15\hsize \epsfbox{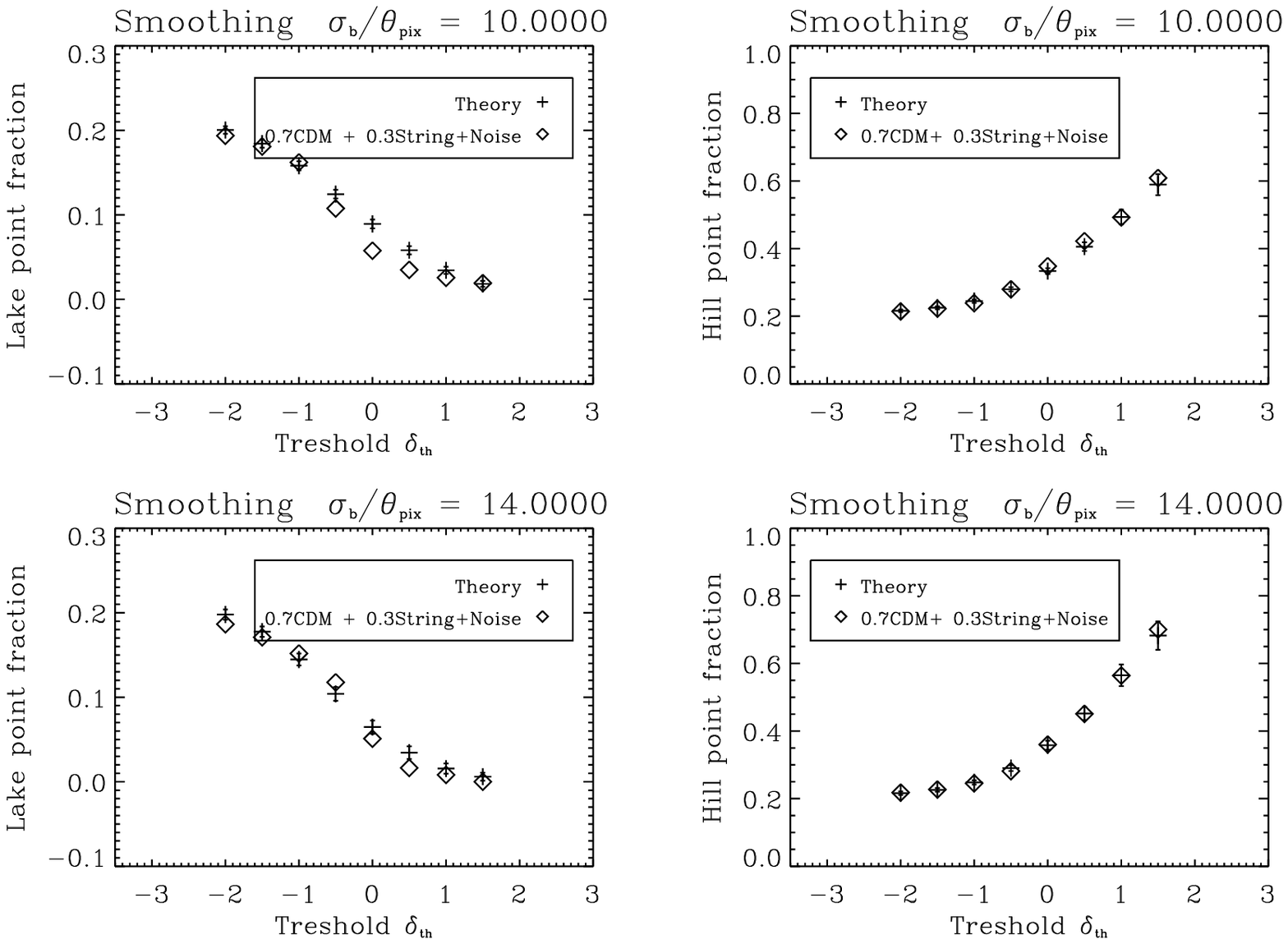}}
\caption{Same as Fig.~\ref{plot_cdmt_topo2_cdm_noP} but for the an
hybrid model with $\beta=0.3$.\label{plot_cdmt_topo_2_st0d3_noP}} 
\centering
\centerline{\epsfxsize=1.15\hsize \epsfbox{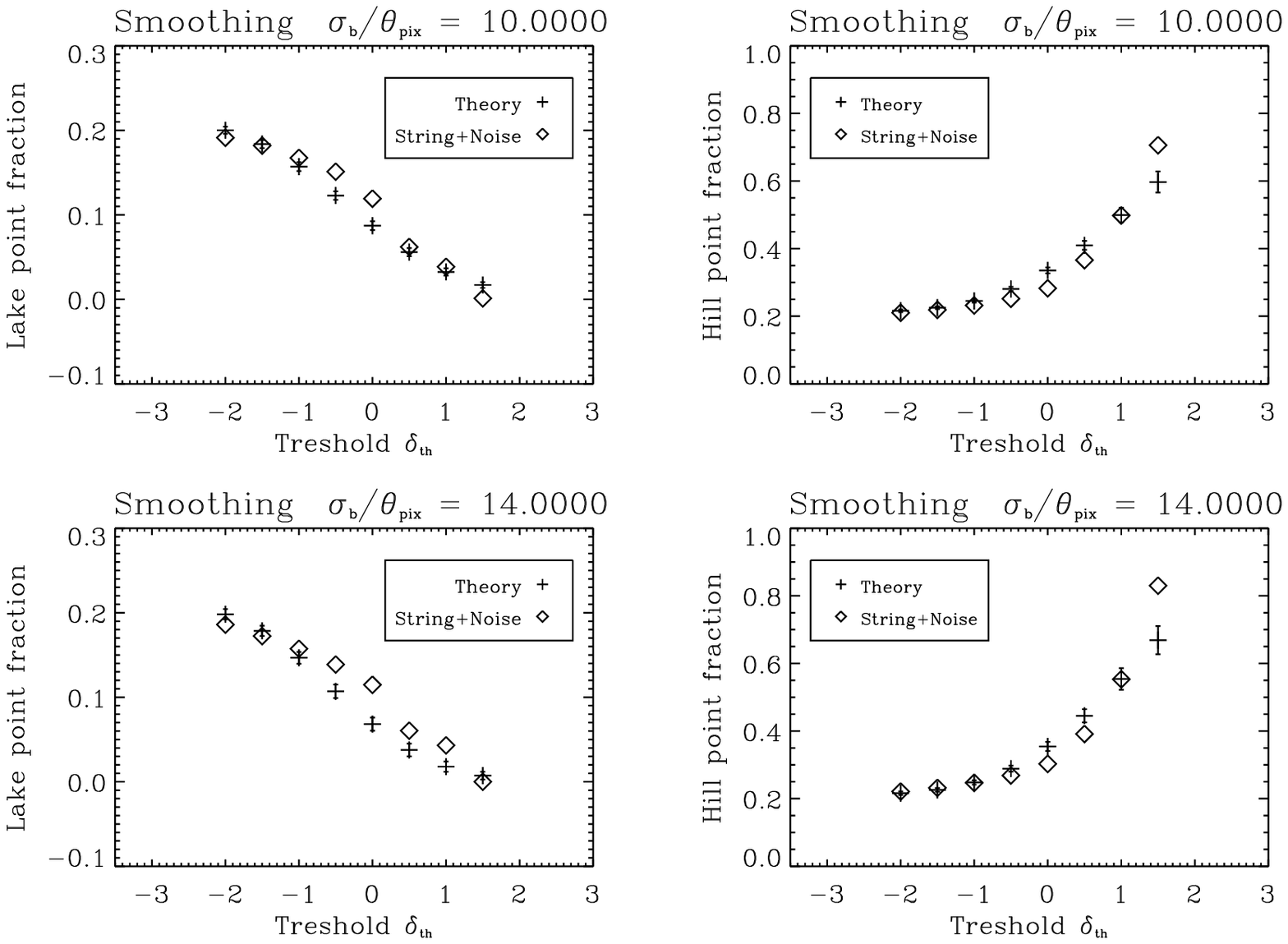}}
\caption{Same as Fig.~\ref{plot_cdmt_topo2_cdm_noP} but for the pure string
model ($\beta=0.0$). \label{plot_cdmt_topo_2_st_noP}}
\end{figure}

\setcounter{figure}{10}
\begin{figure*}
\centering{
\centerline{\epsfxsize=11cm \epsfbox{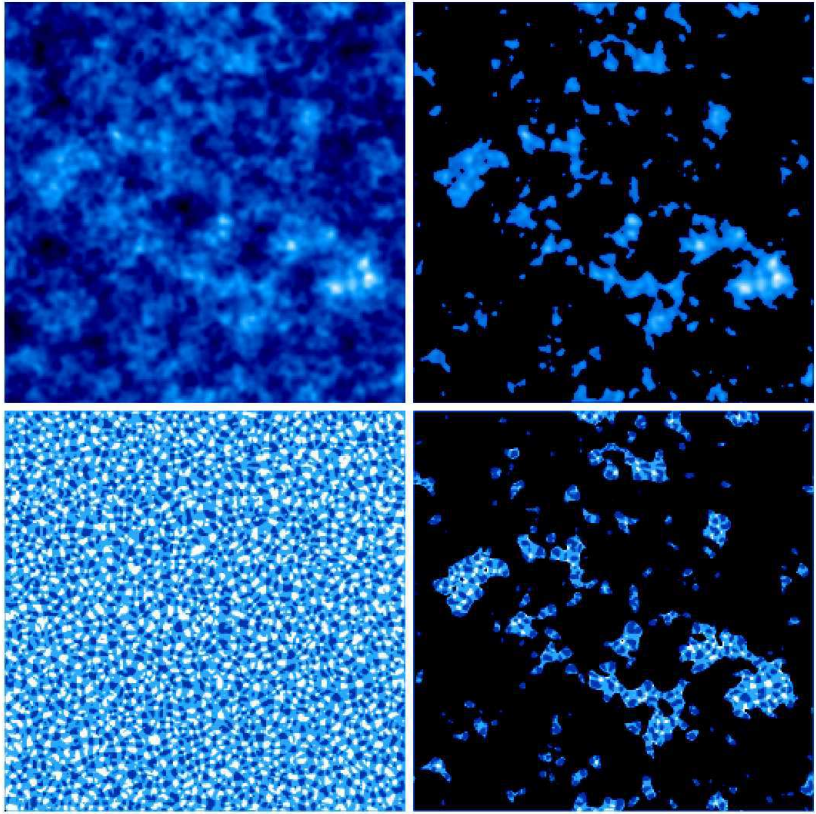}}
\caption{Same as in  Fig.~\ref{panel_cdm} but
for a noise free mixed model with $\beta = 0.7$. Note how difficult
it is to distinguish it from the mere CDM map.\label{panel_mixt}}}
\end{figure*}

In Figs.~\ref{plot_cdmt_topo2_cdm_noP}, \ref{plot_cdmt_topo_2_st0d3_noP} and
\ref{plot_cdmt_topo_2_st_noP}, we display some 
measurements of functions ${\mathcal P}_{\rm lake}(\delta_{\rm th})$ and
${\mathcal P}_{\rm hill}(\delta_{\rm th})$ in the Planck case. Three
values of $\beta$ are considered: pure CDM with $\beta=1$, hybrid
model with $\beta=0.3$ and pure string model with $\beta=0$. Again,
On each figure, the Gaussian limit
obtained from the procedure explained in detail in {\S}~\ref{ng_method} is compared
to the measurements. The error bars on
the theoretical predictions correspond to then the 
$1\sigma$ dispersion over 300
realizations of pure random Gaussian fields with same power-spectrum
and noise properties as the data, as explained in end of {\S}~\ref{ng_method}. 
Very good agreement is found for CDM, as expected,
while the hydrid model
exhibits some discrepancies, especially for ${\cal P}_{\rm
lake}$. These discrepancies are overwhelming in the string only
model. Note that since we fit the measured value of
${\mathcal P}_{\rm hill}(\delta_{\rm th}= 1)$ to estimate
the Gaussian prediction, one expects non Gaussian features to
show up more in the measured shape of function ${\cal P}_{\rm
lake}(\delta_{\rm th})$ than that of function 
${\mathcal P}_{\rm hill}(\delta_{\rm th})$, as indeed seen
at least for the mixed model. Note also
that if for the mixed model, the measured ${\cal P}_{\rm
lake}(\delta_{\rm th})$ tends to lie below the Gaussian prediction,
the reverse happens for the pure string model (this seemingly
counterintuitive result stems from our comparing the data to a
``theoretical'' curve which is different in the 2 cases) . 
Finally, it is important as well to recall that data points plotted on 
Figs.~\ref{plot_cdmt_topo2_cdm_noP}--\ref{plot_cdmt_topo_2_st_noP} 
represent only 2/3 of all the points used for the $\chi^2$
computation since we consider one more smoothing scale,
$\sigma_{\rm b}/\theta_{\rm pix} = 6$. 

To illustrate more qualitatively these measurements, Figures \ref{panel_mixt}
and \ref{panel_st} display, similarly as
in Fig.~\ref{panel_cdm}, an example of initial smoothed {\it noise free}
map, its thresholded counterpart (with $\delta_{\rm th}=1$) and the corresponding
local curvature maps with lake, saddle and hill points.
Figure \ref{panel_mixt} corresponds to the
hybrid model case  ($\beta=0.3$) while Figure~\ref{panel_st}
corresponds to the string only model
($\beta=0.0$). A strong similarity can be seen between 
the mere CDM and the hybrid models despite the differences 
in measured ${\cal P}_{\rm lake}$ functions.
Note also the peculiar patterns in the pure string
curvature map: there seem to be rather extended saddle point regions, 
which calls for other specific pattern detection statistics. 

Table~\ref{table_chi2} summarizes the results obtained for $\chi^2$ using the noise level 
of the best channel of either Planck or MAP. It shows, for various values of $\beta$, 
the measured $\chi^2$, the $\chi^2$ per dof, as well as the
probability (noted Prob.) that a random variable, in a
$\chi^2$-distribution with the relevant (number of)  
dof (shown above to be
adequate) is greater than the obtained $\chi^2$. This quantity gives
therefore the significance of the detection.    

Roughly speaking, the smaller the $\chi^2$, the better is the
agreement with a Gaussian distribution. Reversely, the higher the obtained
$\chi^2$, the more unlikely does the analyzed map follow a  Gaussian distribution.

Note that the numbers displayed in Table~\ref{table_chi2} 
{\it represent in no way} the best Planck and MAP can provide with our method, 
since we only considered 0.3\% of the sky for {\it one single
channel}. Nevertheless, these figures give some trends of the ability
of our method, the purpose of this paper.  

With these limitations in mind, we see that MAP can get a meaningful
detection (1\% level) of the non-Gaussianity induced by string like
background with $\beta = 0.7$, corresponding to a number $1-\alpha =
0.15$,  a number slightly lower than the constrains from Bouchet \etal
(2000), $1-\alpha = 0.18$. Any smaller non-Gaussian 
contribution is not detected with a good significance. On the other
hand, we see that Planck could obtain, using this method, a highly
significant detection level, \ie smaller than 0.1\%, for $\beta$
between 0.8 and 0.9, \ie $0.05 \leq 1-\alpha \leq 0.12$. 


\section{Discussion and Conclusion}
\label{discuss}

\setcounter{figure}{12}
\begin{figure*}
\centering{
\centerline{\epsfxsize=11cm \epsfbox{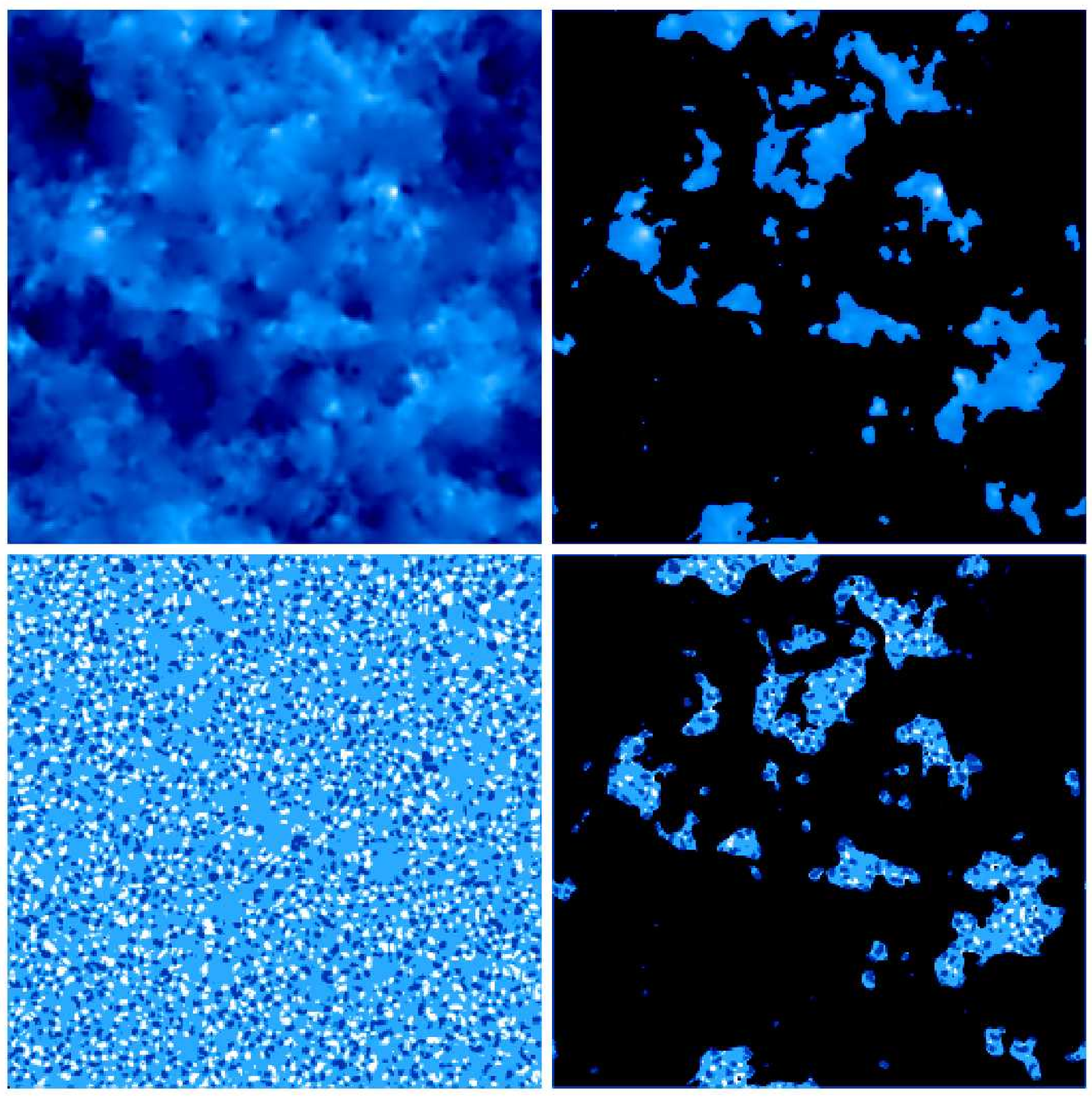}}
\caption{Same as in Fig.~\ref{panel_cdm} but
with noise free string induced only temperature fluctuations. 
Note the interesting structures in the curvature map. \label{panel_st}}}
\end{figure*}

\begin{table*}
\begin{center}
\begin{tabular}{c|ccc|ccc}
\hline 
\hline
\multicolumn{1}{c}{} & \multicolumn{3}{c|}{MAP\quad(dof = 31) } &
\multicolumn{3}{c}{Planck\quad(dof = 47)}\\      
\cline{2-7}
\multicolumn{1}{c}{} & $\chi^2$ & $\chi^2/{\rm dof}$ 
& Prob. &$\chi^2$ & $\chi^2/{\rm dof}$ &\multicolumn{1}{c}{Prob.} \\ 
\cline{1-7}
$\beta = 1.0  $ & 34.1  & 1.1 & 0.32 & 46.6 & 0.99 & 0.49 \\
$\beta = 0.0  $ & 109.3 & 3.52 & $3.1\ 10^{-10}$  & 879.3 & 18.7 & 0.0 \\
\hline
$\beta = 0.7  $ & 120.9 & 3.9 & 4.01 & 217.1 & 4.6 & 0.0 \\
$\beta = 0.8  $ & 68.2 & 2.2  & $1.3\ 10^{-4}$ & 180.5 & 3.84 & 0.0 \\
$\beta = 0.9  $ & 55.8 & 1.8  & $4.0\ 10^{-3}$  & 101.1 & 2.15 & $7.0\ 10^{-6}$ \\
\hline
\hline
\end{tabular}
\caption[ ]{The measured $\chi^2$, $\chi^2/{\rm dof}$
and significance probability (Prob.) are shown for various values
of $\beta$ and for our virtual MAP and Planck experiments, as explained
in the text.}
\end{center}
\label{table_chi2}
\end{table*}

In this paper, we have tested a statistic based on local
curvature measurements by counting three well defined types
of points according to the sign of the eigenvalues of the Hessian
of a temperature map. In the first section, we computed the relevant theoretical
expectations. We validated them and demonstrated our ability 
to properly measure them in the next section and eventually introduced this statistic as an accurate
non-Gaussianity test (by performing a $\chi^2$ like analysis) in the
last sections, where we applied it to the case of mixed models with
realistic noise levels. Even if our physical simulation might be
somewhat simplistic (\eg no pre-lss string contribution, no
foregrounds residuals, etc.), they showed clearly that we are able by using this
technique in the Planck context, to distinguish a non-Gaussian
background amplitude around 
10\% in temperature rms (5\% in $C_{\ell}$)  with a good significance 
using only 0.3\% of the sky and one single channel. Consequently,
these results are definitely positive and, as it is, this method seems
to be ready to be applied on true data.  

Naturally, in practice this test should be used in conjunction with other
more standard statistical measurements, like 2-, 3-, 4-,\ldots points correlation
functions or their harmonic transform, power spectrum, bispectrum,
trispectrum \ldots or even already presented more sophisticated ones like the
distribution of wavelet coefficients, etc. This array of methods
might actually help in identifying the source of
non-gaussianity detected in a realistic context. Indeed, in practical
situations, important 
non-Gaussian effects might be induced by CMB contaminants, in
particular galactic foregrounds, 
residual map stripping or point sources, or any instrumental systematic 
error. 

\subsection{Possible advantages}

One important advantage of these statistic is its local nature, \ie
the fact that we are only interested by the 8 neighboring
pixels. The sphericity of the sky is thus not an issue for this method
and so is the pixelisation, provided that the smoothing scale is about
3 times the pixel size\footnote{Which is convenient, since
pixelisation of low noise experiments are usually done at a size
similar to one-third of the FWHM of the beam.} (see {\S}\ref{gamma_eff}). 

A second advantage of the method lies in the fact that it is
mathematically well-defined and simple since it depends only on one
single spectral parameter, $\gamma$ that can be either measured from
the fractions themselves or from the $C_{\ell}$ measurement (see
{\S}\ref{gamma_eff}). Thus measurements are well understood and well
controlled from a theoretical point of view.

\subsection{Outlooks}

As we mentioned earlier, by moving to second order derivatives, this
work opens the way to many more characterization of random Gaussian
field. We expect that an analogous work could be  done by considering
critical points only, which would lead, \eg, to a measurement  of the
Euler characteristic as  a function of threshold. However, we expect
this restriction to be less efficient in probing non-Gaussianity since
it appears that by keeping all types of points we benefit more from
the full structure of the probability distribution function. This
point will be demonstrated in a future work. In addition, before applying the
method to real data, it will also be interesting 
to compare it's ability to that of other estimators on the same test
protocole, but extended to include other families of models, as \eg \cite{LyWa01}. 

\section*{Acknowledgments}

We are grateful to Roland Triay for organizing a prolific ``Cosmological
school'' at Porquerolles where this work has been initiated, to
Francis Bernardeau for useful discussions at an early stage of this work
and to Christophe Pichon for valuable help on the use of
Mathematica and keyboards. S.C. thanks Jean-Philippe Belial for quite
useful suggestions on the methodological approach. We are grateful to
J{\'e}r{\^o}me Martin and Patrick Peter for a careful reading of the manuscript.  


\end{document}